\documentclass[conference]{IEEEtran}
\IEEEoverridecommandlockouts
\usepackage{cite}
\usepackage{amsmath,amssymb,amsfonts,multirow}
\usepackage{algorithmic}
\usepackage{graphicx}
\usepackage{textcomp}
\usepackage{xcolor}
\def\BibTeX{{\rm B\kern-.05em{\sc i\kern-.025em b}\kern-.08em
    T\kern-.1667em\lower.7ex\hbox{E}\kern-.125emX}}
\begin{document}

\title{Optimizing Speech Emotion Recognition using Manta-Ray Based Feature Selection}

\author{\IEEEauthorblockN{Soham Chattopadhyay}
\IEEEauthorblockA{\textit{Dept. of Electrical Engineering} \\
\textit{Jadavpur University}\\
Kolkata, India \\
chattopadhyaysoham99@gmail.com}
\and

\IEEEauthorblockN{Arijit Dey}
\IEEEauthorblockA{\textit{Dept. of Computer Science \& Engineering} \\
\textit{Maulana Abul Kalam Azad University of Technology}\\
Kolkata, India \\
arijjitdey3413@gmail.com} 

\and

\IEEEauthorblockN{Hritam Basak}
\IEEEauthorblockA{\thanks{$^*$ Corresponding author}
\textit{Dept. of Electrical Engineering} \\
\textit{Jadavpur University}\\
Kolkata, India \\
hritambasak@gmail.com}
}

\maketitle

\begin{abstract}
Emotion recognition from audio signals has been regarded as a challenging task in signal processing as it can be considered as a collection of static and dynamic classification tasks. Recognition of emotions from speech data has been heavily relied upon end-to-end feature extraction and classification using machine learning models, though the absence of feature selection and optimization have restrained the performance of these methods. Recent studies have shown that Mel Frequency Cepstral Coefficients (MFCC) have been emerged as one of the most relied feature extraction methods, though it circumscribes the accuracy of classification with a very small feature dimension. In this paper, we propose that the concatenation of features, extracted by using different existing feature extraction methods can not only boost the classification accuracy but also expands the possibility of efficient feature selection. We have used Linear Predictive Coding (LPC) apart from the MFCC feature extraction method, before feature merging. Besides, we have performed a novel application of Manta Ray optimization in speech emotion recognition tasks that resulted in a state-of-the-art result in this field. We have evaluated the performance of our model using SAVEE and Emo-DB, two publicly available datasets. Our proposed method outperformed all the existing methods in speech emotion analysis and resulted in a decent result in these two datasets with a classification accuracy of 97.49\% and 97.68\% respectively.

\end{abstract}


\section{Introduction}
Expressing emotions is an important part of communication among human and non-human primates and it is the most common way to express love, sorrow, anger, hatred, or any other state of mind [1], [6]. Even, non-speaking living beings also find their ways to express their emotions. We find the verbal communication and associated emotions so important that we often miss the scarcity of those in text messages or emails and hence switch to the usage of emojis. Since emotions help us to understand each other better, we have tried to implement emotion recognition using computers as well. With the recent advancements in natural language processing and speech to text conversions, scientists have often looked upon the automation of information parsing from speech audio and artificial intelligence has been successfully deployed for generating auto-replies; the chat-bots and recent speaking-humanoid robots are exemplary evidence of these advancements. However, the emotion analysis of auditory signals has been studied lately by researchers and several improvements have been made in this domain for the last two decades. 
However, in real life it is a difficult work to predict human emotion from a conversation. The most important work in the acoustic signal processing field is to extract features properly. Now a days different machine learning and deep learning models are made to deal with signal processing task. In this literature we have used different methods for feature extraction task. However, every extracted features from the audio is not accurate and can be redundant. That is the main reason researchers find it difficult to remove redundancy from the feature vector. There are different features of a audio signal such that, 1. Temporal features and spectral features. We are mainly focusing on temporal features and there are some standard techniques, Mel-frequency cepstral coefficient(MFCC) [39],  Linear prediction coefficients (LPC) [40], Linear prediction cepstral coefficients (LPCC) [41], Perceptual Linear Prediction (PLP) [42] etc. Combination of these features give the accurate result and can explain the nature of the audio signal properly.

 Although the Speech Emotion Recognition (SER) has versatile applications [11], there exists no generalized or common consensus on categorization and classification of emotions from speech signals as the emotions are subjective property of human. Besides, the emotions may vary in intensity, mode of expression from person to person, and may vary or frequently misinterpreted by people. Therefore, the automatic AI-based classification of emotion from auditory signals is the test-bed of assessing the performance of several existing feature extraction and classification methods [14]. Though the discrete and dimensional models have shown improved performance in recent times, there is a huge scope of improvements as it remains an open problem in a bird’s eye view.

\begin{figure*}
\centerline{\includegraphics[width=1.8\columnwidth]{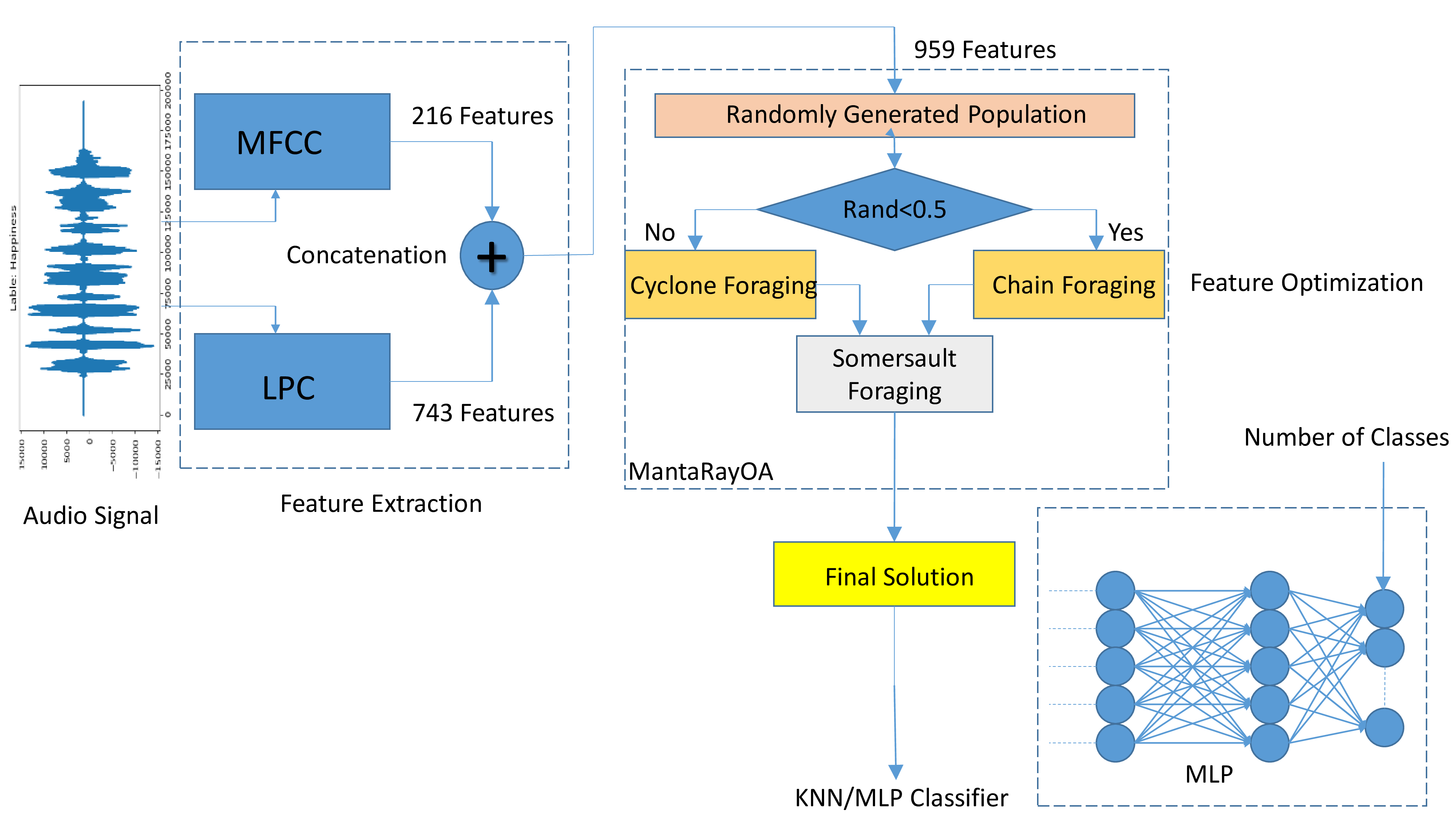}}
\caption{Workflow diagram of our proposed method.}
\label{fig1}
\end{figure*}

The human voice can have features from different modalities, though the most predominant ones are: (1) voice quality, (2) Teager energy operator, (3) prosodic, and (4) spectral features, [8] though, the classifier performance can be improved by incorporating features from other different modalities too. As supervised learning is based on the feature quality and accurately labeled dataset, the performance of these classification problems has highly relied upon the efficacy and experience of the feature engineer performing the feature extraction as more imagination can open up the possibility of new multimodal feature extraction. But this requires lots of time and imagination and still, it is impossible to extract all the important, high-quality, and insightful features from auditory signals. 

\textbf{Contribution of this paper:} The contribution of this paper can, therefore, be considered as a two-fold contribution: 
First, we explore the feature concatenation in auditory signal processing. Merging of features from different sources has been successfully utilized in the image processing task before [1-4]. Being inspired by these, we have successfully evolved our proposed approach based on feature concatenation. Secondly, we have proposed a bio-inspired meta-heuristic Manta-Ray foraging Optimization algorithm for feature selection and removal of redundant features. This algorithm was evolved in 2020 by Zhao \emph{et al.} [5], being inspired by the behavioral study of an aquatic species named Manta Ray.  

Our proposed method outperformed all the existing methods in speech emotion analysis and resulted in a state-of-the-art result in the SAVEE dataset with a classification accuracy of 97.49\% on average. We also validated our method on the Emo-DB dataset that also produced a state-of-the-art result with a classification accuracy of 97.68\%. Fig 1. shows the workflow digram of our proposed method.

 The results of our experiment are described in Section 4. 

\section{Related Works}
Speech emotion recognition deals with the basic task of identification of emotions and expressions from audible sounds and an intensive research in this domain started in early 90’s. However, recent advancements in technology and computational resources have made this study diverse and compelling. Following are the three major approaches made by the scientists recently to address this problem.
\subsection{Machine learning algorithms}
The Research in this field started in the early 1990's, whereas the first significant result was obtained by Nakatsu \emph{et al.} [7] in 1999 where they had used speech power and basic Linear Predictive Coding (LPC) features from 100 human specimens, equally distributed among males and females. The simplest neural network model they used, produced a result of 50\% recognition accuracy. Later, Schuller \emph{et al.} [9] proposed a hidden-Markov model, and validation was done on the Emo-DB and VAM dataset that produced correct classification accuracy of 76.1\% and 71.8\% respectively, using raw contours of ZCR, energy and pitch features. Later, the improved version of their proposed Markov model produced a state-of-the-art result on German and English speech signals containing 5250 samples with an average accuracy of 86.8\% with global prosodic features using continuous HMM classifier and pitch and energy-based features [28]. Rao \emph{et al.} [11] used the Support Vector Machine classifier using RBF kernel with approximately 67\% classification accuracy using prosodic features. LFPC features were extracted and the HMM classifier was used for classification of Mandarin language by [12] with an average precision of 78.5\%. Another important result was produced by Wu \emph{et al.} [13] by using an SVM classifier that reached a classification accuracy of 91.3\% on the Emo-DB dataset and 86\% accuracy on the VAM dataset by using prosodic, ZCR and TEO features. Deng \emph{et al.} [6] validated their proposed method on four different datasets: Emo-DB, VAM, AVIC, and SUSAS by using autoencoder classifiers. The features used were MFCC and other LLD features (example: ZCR and pitch frequency etc.) and denoising was performed before the classification task and presented classification accuracy of 58.3\%, 60.2\%, 63.1\%, and 58.9\% respectively.  
\subsection{Deep learning methods}
 With the recent advancement in deep learning in the last few years, scientists have tried to exploit the ability of Deep Neural Networks models to learn high-quality semantic information and invariant features from different types of datasets [14, 15]. A few recent studies provided results that supported this conclusion that DNNs are equally efficient and useful for speech-emotion classification. Rong \emph{et al.} [10] had used KNN classifiers for classifying the Mandarin dataset using ZCR, spectral, and energy features. Stuhlsatz \emph{et al.} [16] and Kim \emph{et al.} [17] used the utterance level features to train the DNN models. Rozgic \emph{et al.} trained the DNNs using the combination of lexical and acoustic features. Unlike these existing DNN models, that directly used acoustic features learned from sources, our proposed method uses optimization in between for the improvement in performance.  
 \subsection{Feature selection and optimization}
 Research works have also been made on selection of optimal features from the huge feature set, extracted from speech data using various feature extractors. Feature selection is a crucial step in system development to identify emotions in speech. Recently, the interaction between features collected from the auditory signals was rarely considered, which may result to redundant features and sub-optimal classification results. Gharavian et al.[37] used FCBF feature selection methods and Genetic algorithm based ARTMAP neural networks for optimization and classification. Particle Swarm Optimization (PSO) is another popular optimization algorithm, used for feature selection in SER by Muthusamy et al.[38]. Both of them reported a high accuracy as compared to the previous works on various datasets. These laid the groundwork for the analysis of different feature selection algorithm empirically which was further taken forward by others. Liu et al.[33] used optimal features before feeding them to the extreme learning machine (ELM) for classification whereas Ververidis et al.[34] suggested sequential floating forward feature selection (SFFS) so that the features obey the multivariate Gaussian distribution before using Bayesian classifier for classification. Sheikhan et al.[36] proposed ANOVA feature selection method in SER task and used machine learning based modular neural-SVM classifier. Özseven [35] suggested a novel feature selection algorithm and suggested the superiority of that by validating the results on some standard datasets. These works helped us to further investigate in the SER feature selection task prior to classification.

\section{Materials and Methods}
This section describes the workflow of this experiment which consists of the following steps: (1) Data acquisition and preprocessing, (2) Feature Extraction, (3) Feature selection using optimization algorithm, and (4) Classification and analysis of results. 

\subsection{Data acquisition} The initial step of the experiment is dataset collection and preprocessing. We used two publicly available datasets for this experimentation purpose. These are: 

\paragraph{SAVEE dataset} The Surrey Audio-Visual Expressed Emotion (SAVEE) Database consists of speech recordings from four British actors consisting of 480 samples in total, collected, and labeled with extreme precision and by using high-quality equipment. The dataset is classified into seven different emotional categories: happiness, sadness, disgust/fear, neutral, common, anger, and surprise. The sentences were chosen by experts carefully from TIMIT Corpus and were balanced phenotypically in every category.   

\paragraph{Emo-DB dataset} The Berlin Emo-DB dataset contains emotion speech data from 10 different speakers and contains 500 labeled samples in total. It also contains 7 categories of emotion-speeches: normal, anger, sadness, happiness, disgust, anxiety, and fear.  

\subsection{Preprocessing}
For every kind of signal processing task, the pre-processing of sample data plays a vital role in determining the performance of a model. Some simple audio pre-processing techniques which have improved the exploitation of our model are discussed below:
\paragraph{Pre-emphaize}Pre-emphasis step is mainly carried out as it synthesizes the normal form of any amplitude signal. The main idea behind this is to flatten the speech spectrum, which can be done by implementing a high-pass Finite Impulse Response (FIR) filter. The expression of the filter in a discrete frequency domain is given by:
\begin{equation}
F(z)=1-Az^{-1}
\end{equation}

To normalize the signal, firstly the maximum value of the signal has to be taken as the nominator and dividing the signal with it. Thus the entire signal is normalized between -1 and 1. For smooth transactions between frames 50\% overlaps of consecutive frames are accepted. The mechanical acoustic signal can be stable in the range of 50ms to 200ms and so we have selected a short window for better feature extraction.
 \paragraph{Framing}For further processing, the signal is divided into small frames such that to get a sequence of frames forming the entire original signal. This is done so that a long signal can be analyzed independently in small frames and can be expressed through a single feature vector. Some aspects of framing like frameshift are the time difference between two starting points of two consecutive frames and frame length is the duration of time for each frame. 

\paragraph{Windowing}For audio signals it is quite common of having discontinuities at the frame edges of the signal. This phenomenon often causes bad performance in the audio processing task. To get rid of this some trapped windows such as the Hamming window is applied at each frame. The general expression of the Hamming window is:
\begin{equation}
W\textsubscript{h}=a-b\times cos\left(\frac{2\pi n}{N-1}\right)
\end{equation}
where \emph{a=0.54, b=0.46}, and \emph{N} is the number of samples in a partition of the data into some random complementary subsets.

\subsection{Feature extraction}
Features play an important role in any classification task as all the information of speech data is embedded in these features. Therefore, high-quality and accurately extracted features may contribute immensely towards the improved performance of a classifier. In this experiment, we have extracted two short-time features from the dataset: Mel Frequency Cepstral Coefficient (MFCC) and the Linear Predictive Coding (LPC). The brief description of the feature-extraction is described below.
\paragraph{MFCC features}
MFCC features are based on the human auditory sensation characteristics. The mathematical simulation of hearing is done in MFCC by using some non-linear frequency units. We normally use the Fast Fourier Transform or Discrete Fourier transform for the conversion of acoustic signals from the time domain to the frequency domain for each sample frame. The Fast Fourier Transform (FFT) is described by the following equation:
\begin{equation}
y(k)=\sum\limits_{p=0}^{N-1} y[n]. e^{-j\frac{2\pi pk}{N}}, \;\; 0\leq k \leq (N-1) 
\end{equation}
where \emph{y[n]} represents the signal in the time domain and the domain-converted signal \emph{y(k)} is in the frequency domain and N is the number of samples in every frame. Next, we calculate the Disperse Power Spectrum by using the following equation:
\begin{equation}
Power \; Spectrum(PS\textsubscript{P})=y(k) \cdot \land y(k)
\end{equation}

After this, we have the MEL spectrum by the \emph{PS\textsubscript{p}} in the triangular filter bank. The filter bank consists of the series of triangular filters with the cutoff frequency determined by the center-frequencies of the two presently adjacent filters. These filters are linear in MEL frequency coordinates. The scale is equivalent to the span of every filter and the value for the span is set to \emph{167.859} in this project. The frequency response of the triangular filter can be calculated as:
\begin{equation}
\text{F}[n]=\left\{
\begin{array}{lr}
0, \;n<g(p-1)\\
\frac{2(n-g(p-1))}{(g(p+1)-g(p-1))(g(p)-g(p-1))},\;g(p-1)\leq n \\
\frac{2g((p+1)-n)}{(g(p+1)-g(p-1))(g(p+1)-g(p))}, \; g(p)\leq n\\
0, \; n\geq g(p+1)
\end{array}
\right\}
\end{equation}
where \emph{p=1,2,.....,12}, \emph{g(p)} is the centre frequency of the filter, \emph{n=1,2,....,(N/2-1)} where \emph{N} being number of samples per frame.
To improve the quality of the features we used logarithmic spectrum of power spectrum on the signal and which is represented by the following equation. 
\begin{equation}
\text{L}(p)=ln\left( \sum\limits_{n=0}^{N-1} |PS\textsubscript{p}|^2 \text{F}[n]\right),\;0\leq p\leq N
\end{equation}
where L\emph{(p)} is the logarithmic spectrum, F\emph{[n]} is the series of filters, \emph{PS\textsubscript{p}} is defined earlier, \emph{N} is the number of samples per frame.

Finally, the Discrete Cosine Transform (DCT) of the logarithmic spectrum of the filter banks are calculated that gave the MFCC feature which is described in the following equation.
\begin{equation}
\text{F}[n]=\sum\limits_{p=1}^{N-1}\text{L}(p)cos\left(\frac{(M-1)\times k \times \pi}{2N}\right), \;0 \leq k \leq N
\end{equation}

\paragraph{LPC features}
The speech signal is sequential data and so let’s assume the voice acoustics of \emph{n\textsuperscript{th}} speech sample is \emph{P[n]} which can be shown as the combination of previous \emph{k} samples. The \emph{n\textsuperscript{th}} speech sample can be mathematically represented as:
\begin{equation}
\text{S}[k]=\sum\limits_{j=1}^M a\textsubscript{j} S[k-j]+ H\times E[k]
\end{equation}
where S\emph{[k-j]}, where \emph{j = 1, 2, 3 . . . k}, is the j\textsuperscript{th} test sample, \emph{H} stands for gain factor, \emph{E[k]} denotes the excitation of the k\textsuperscript{th} sample, \emph{a\textsubscript{j}} is the vocal tract vector coefficient. LPC is also known as inverse filtering because it determines all zero filters which are also inverse of the vocal tract model. 

\subsection{Manta Ray Optimization}In this work we have used the Manta-Ray Foraging Optimization (MRFO) algorithm for feature selection. It is a bio-inspired optimization algorithm that works on three unique foraging strategies of Manta Ray. 

\paragraph{Chain foraging} Manta Rays have their unique approach towards their prey, which is imitated in our project. They tend to swim towards the highest concentration of planktons, i.e. the best possible position of food. In our algorithm, we set some random initial values and ask them to move towards the optimal solution in every iteration. Though the actual optimal solution is not known, this algorithm assumes that the best solution is the one where the plankton concentration is highest. All the members of the group proceed towards the plankton concentration, by following the previous member of the group, except the first one. This is known as forming a foraging chain, which is imitated in our experimentation. Fig. 2 is a simulation diagram of chain foraging in 2-D space. Mathematically, the chain foraging is represented by:
\begin{figure}
\centerline{\includegraphics[width=\columnwidth]{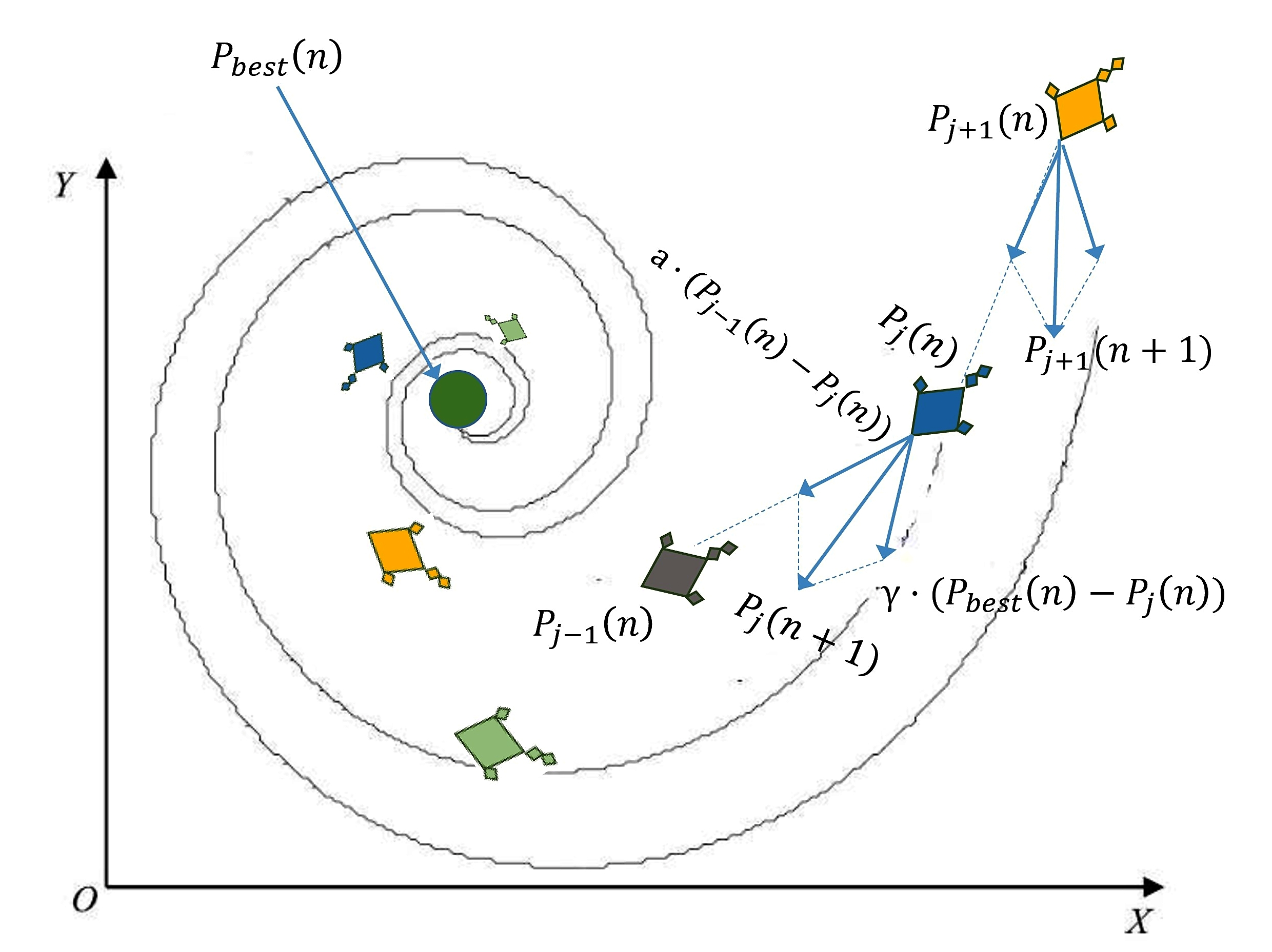}}
\caption{Simulation model of chain-foraging of Manta Ray in two-dimensional space}
\label{fig2}
\end{figure}
 
\begin{equation}
P\textsubscript{j}\textsuperscript{k}(n+1)=\left\{
\begin{array}{lr}
P\textsubscript{j}\textsuperscript{k}(n)+a\left(P\textsubscript{best}\textsuperscript{k}(n)-P\textsubscript{j}\textsuperscript{k}(n)\right)+\\b\left(P\textsubscript{best}\textsuperscript{k}(n)-P\textsubscript{j}\textsuperscript{k}(n)\right);\;j=1\\
\\
P\textsubscript{j}\textsuperscript{k}(n)+a\left(P\textsubscript{j-1}\textsuperscript{k}(n)-P\textsubscript{j}\textsuperscript{k}(n)\right)+\\b\left(P\textsubscript{best}\textsuperscript{k}(n)-P\textsubscript{j}\textsuperscript{k}(n)\right);\;j=2,3,..,N
\end{array}
\right\}
\end{equation}
 \begin{equation}
 b=2a\sqrt{|ln(a)|}
 \end{equation}
where \emph{P\textsubscript{j}\textsuperscript{k}(n)} is the position of a j\textsuperscript{th} agent in k\textsuperscript{th} dimension, \emph{a} is the random vector in closed range in 0 to 1, \emph{b} is known as weight coefficient, \emph{P\textsubscript{best}\textsuperscript{k}(n)} is the best plankton concentration position. The position of the (j+1)\textsuperscript{th} is determined by the previous j agents.

\paragraph{Cyclone foraging} When a group of manta rays gets to know about a high concentration of plankton in the water, they form a chain-like structure that looks like a cyclone and headed towards the plankton. Each of the agents follows the previous agent towards the prey and forms cyclone foraging. Every manta ray doesn’t only form the spiral but also follows the same path and moving towards one step to the plankton following the one in front of it. The mathematical expression for two-dimensional cyclone foraging is given below. 
\begin{equation}
\left\{
\begin{array}{lr}
X\textsubscript{j}(n+1)=X\textsubscript{best}+a\left(X\textsubscript{j-1}(n)-X\textsubscript{j}(n)\right)+\\(X\textsubscript{best}-X\textsubscript{j}(n))e^{cz}cos(2\pi z)\\
\\
Y\textsubscript{j}(n+1)=Y\textsubscript{best}+a\left(Y\textsubscript{j-1}(n)-Y\textsubscript{j}(n)\right)+\\(Y\textsubscript{best}-Y\textsubscript{j}(n))e^{cz}cos(2\pi z)\\
\end{array}
\right\}
\end{equation}
where \emph{z} is a random value in closed range between 0 and 1. This kind of spiral foraging algorithm was also explained by Mirajmili \emph{et al.} in 2016 for Grey Wolf Optimization, however, this is different from that. Fig. 3 is a simulation diagram of cyclone foraging in 2-D space. 
\begin{figure}
\centerline{\includegraphics[width=\columnwidth]{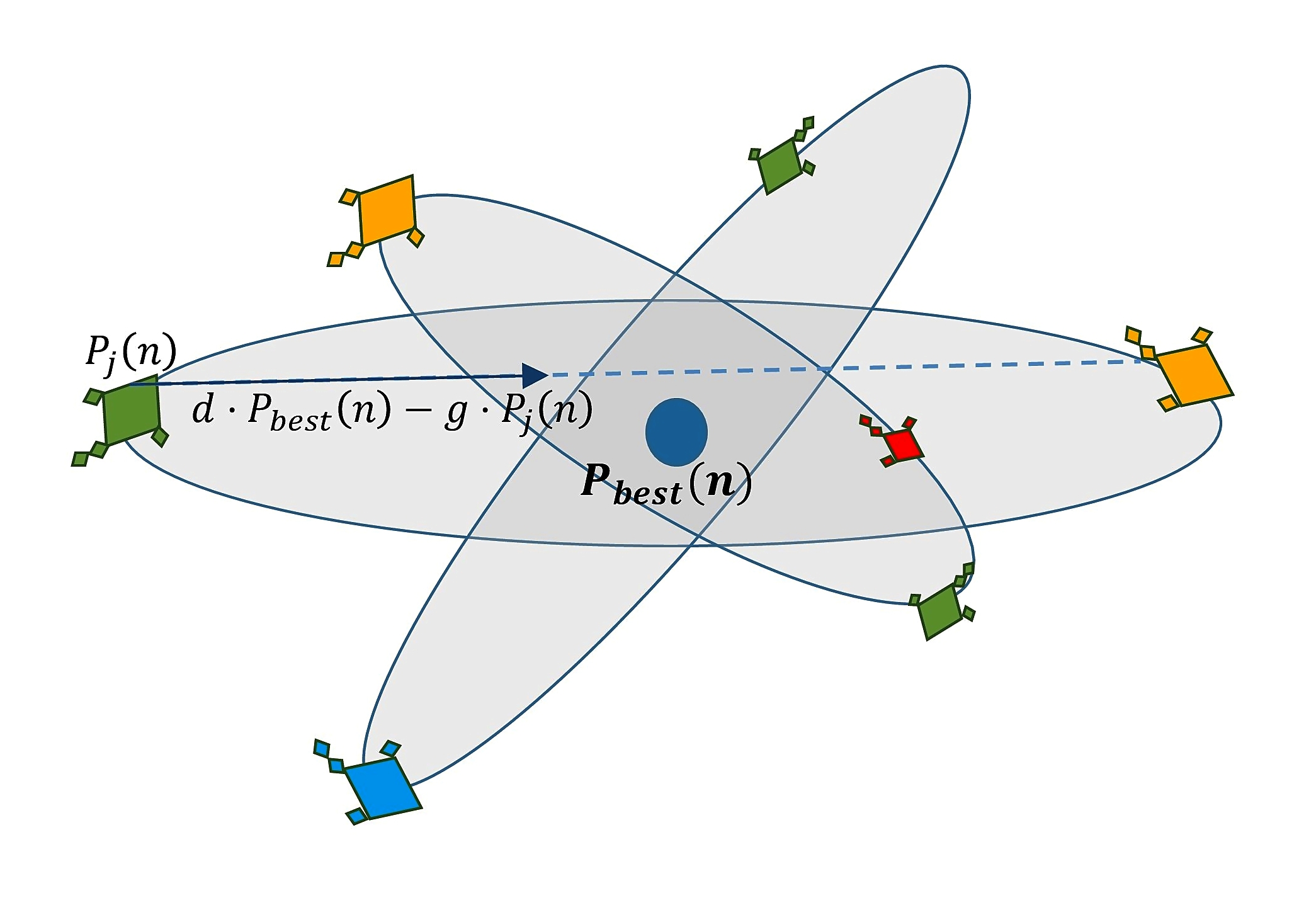}}
\caption{Simulation model of cyclone-foraging of Manta Ray in two dimensional space}
\label{fig3}
\end{figure}
The N-dimensional form of the equation is as followed:
\begin{equation}
P\textsubscript{j}\textsuperscript{k}(n+1)=\left\{
\begin{array}{lr}
P\textsubscript{best}\textsuperscript{k}+a\left(P\textsubscript{best}\textsuperscript{k}(n)-P\textsubscript{j}\textsuperscript{k}(n)\right)\\
+\gamma \left( P\textsubscript{best}\textsuperscript{k}(n)-P\textsubscript{j}\textsuperscript{k}(n) \right);\;j=1\\
\\
P\textsubscript{best}\textsuperscript{k}+a\left(P\textsubscript{j-1}\textsuperscript{k}(n)-P\textsubscript{j}\textsuperscript{k}(n)\right)\\
+\gamma \left( P\textsubscript{best}\textsuperscript{k}(n)-P\textsubscript{j}\textsuperscript{k}(n) \right);\;j=2,3,..,N

\end{array}
\right\}
\end{equation}

 \begin{equation}
 \gamma=2e^{c\frac{I-n+1}{I}sin(2\pi c)}
 \end{equation}
where $\gamma$ is the coefficient of weight, \emph{I} being the maximum iteration and c being a random parameter having value in [0,1].
Each of the search agent perform independent exploration between its current position and the position of the prey. Therefore, this algorithm can efficiently find a best solution in this range. However, we can force any agent to take a new position which is far from its current position by the following equation:
\begin{equation}
P\textsubscript{r}\textsuperscript{k}=B\textsubscript{Low}\textsuperscript{k}+a\left(B\textsubscript{Upper}\textsuperscript{k}-B\textsubscript{Low}\textsuperscript{k}\right)
\end{equation}
 where, \emph{P\textsubscript{r}\textsuperscript{k}} is the newly specified random position of the k\textsuperscript{th} in the N-dimensional space.  \emph{B\textsubscript{Low}\textsuperscript{k}} and \emph{B\textsubscript{Upper}\textsuperscript{k}} are the lower and upper bounds respectively of the N-dimensional space.Thus this algorithm is suitable for finding any best solution in this N-dimensional space.
\begin{figure}
\centerline{\includegraphics[width=\columnwidth]{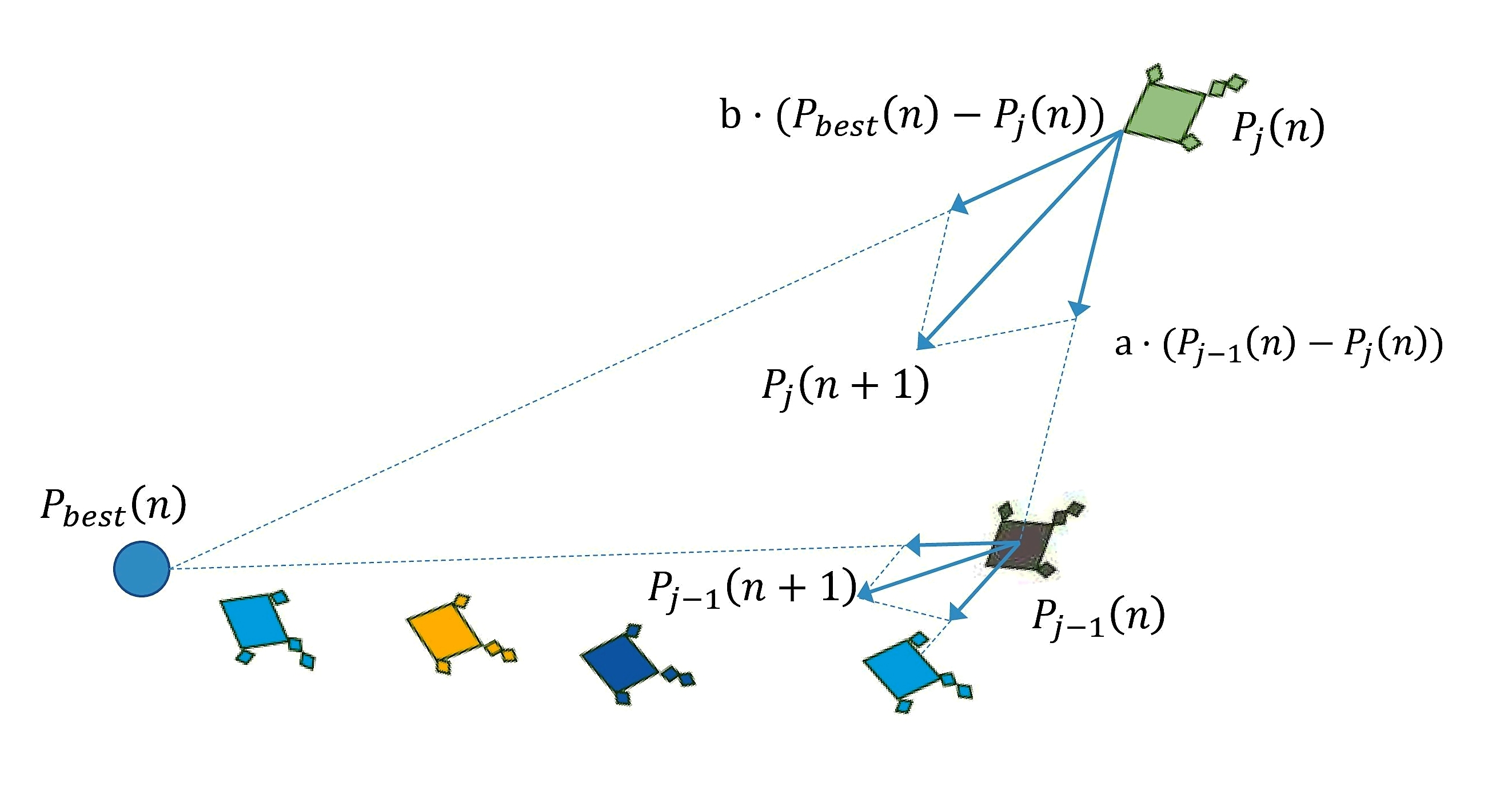}}
\caption{Simulation diagram of somersault-foraging of Manta Ray in two dimensional space.}
\label{fig4}
\end{figure}

\paragraph{Somersault foraging} In this type of foraging the best concentration plankton is considered as a pivot point and all manta rays are moving to the point and eventually update their position around the position of the high plankton concentration(i.e. best solution). Somersault foraging is represented by the equation below:

\begin{equation}
P\textsubscript{j}\textsuperscript{k}(n+1)=P\textsubscript{j}\textsuperscript{k}(n)+F\left(d P\textsubscript{best}\textsuperscript{k}-g P\textsubscript{j}\textsuperscript{k}(n)\right)\;j=1,2,...,N
\end{equation}
where \emph{F} is known as Factor of somersault, \emph{d} and \emph{g} are random parameters in the closed range between 0 and 1. As the equation suggests, this algorithm allows the search agents to update their position at any possible position in the range of its current position and the position of the highest concentration of planktons. With the reduction in the distance between the agents and the optimal solution, the perturbation of the current search agent position also gets decreased. Fig. 4 is a representative diagram of somersault foraging of Manta Ray in 2-D space.

\subsection{Classification} In this current work we have used two different classifiers, one is K-Nearest Neighbourhood classifier and the other one is Multilayer Perceptron classifier commonly known as KNN and MLP classifiers respectively.
\paragraph{KNN Classifier} KNN is a type of supervised Machine Learning model, which finds the ‘feature similarity’ between different data points to prognosticate new data point with a value based on how closely it matches with the other data points in the training dataset. The K-Nearest Neighbour algorithm has some inflexible properties like it is very lazy to learn in type. The fact that it has no proper phase of training and classifies with the entire training dataset as a whole, justifies its lazy learning very clearly. 

To implement this algorithm we have to firstly choose a certain value of ‘K’ that is the number of nearest neighbor points. Thereafter using anyone of the ‘Euclidean’ or ‘Hamming Distance’ or ‘Manhattan’ methods the distance between the test data and each row of training is calculated. Then the points are sorted based on their distance values and top ‘K’ values are chosen amongst them. 
\paragraph{MLP Classifier}Multilayer Perceptron (MLP), colloquially known as ‘Vanilla Network’ is a type of feed-forward Artificial Neural Network. MLP uses back-propagation (a supervised learning technique) to train its weights and biases of different layers. In recent past frequently used activation functions of MLP were sigmoid or tanh, but now as ReLU or Leaky ReLU has proved their better performance so they are used mostly now but specifically in the classification layer still sigmoid or softmax activation functions are used. In this algorithm, changes of each weight in each layer are done by a certain technique called Gradient Descent, given by the following formula:

\begin{equation}
\Delta \beta\textsubscript{ji}(n)=-\alpha\frac{\partial \Omega (n)}{\partial v\textsubscript{j}(n)}y\textsubscript{i}(n)
\end{equation}

Where \emph{$\Omega (n)=0.5\times \sum\limits_{j}l\textsubscript{j}^2(n)$} is the correction that minimizes the prediction error which is calculated by \emph{l\textsubscript{j}(n)=p\textsubscript{j}(n)-y\textsubscript{j} (n)}. Here  \emph{l\textsubscript{j}(n)} is the loss, \emph{p\textsubscript{j}(n)} and \emph{y\textsubscript{j}(n)} are the prediction made by the classifier respectively of the actual class of n\textsuperscript{th} data point of j\textsuperscript{th} node and \emph{$\alpha$} is the learning rate. Now say \emph{$\sigma$'} is the derivative of the activation function. Then for the output node, the derivative would be:
\begin{equation}
-\frac{\partial \Omega (n)}{\partial v\textsubscript{j}(n)}=l\textsubscript{j}(n)\sigma \textsuperscript{,}(v\textsubscript{j}(n))
\end{equation}
For the hidden nodes, the derivative of change follows comparatively complex mathematics given by 
\begin{equation}
-\frac{\partial \Omega (n)}{\partial v\textsubscript{j}(n)}=\sigma\textsuperscript{,}(v\textsubscript{j}(n))\sum\limits_{k}-\frac{\partial \Omega (n)}{\partial v\textsubscript{k}(n)}\Delta\beta\textsubscript{kj}(n)
\end{equation}

\section{Results and Discussions}
The classification of different audio signals is mainly done by the feed-forward neural classifier, specifically the Multilayer Perceptron (MLP).The quantitative measurements are done by using the following equations:

\begin{equation}
    Accuracy\textsubscript{i}=\frac{\sum\limits_{i} M\textsubscript{ii}}{\sum\limits_{i} \sum\limits_{j} M\textsubscript{ij}}
\end{equation}
\begin{equation}
    Precision\textsubscript{i}=\frac{M\textsubscript{ii}}{\sum\limits_{j} M\textsubscript{ji}}
\end{equation}
\begin{equation}
    Recall\textsubscript{i}=\frac{M\textsubscript{ii}}{\sum\limits_{j} M\textsubscript{ij}}
\end{equation}
\begin{equation}
    F1 score\textsubscript{i}=\frac{2}{\frac{1}{Precision\textsubscript{i}}+\frac{1}{Recall\textsubscript{i}}}
\end{equation}
Where M\textsubscript{ij}= the weighted element of the confusion matrix at i\textsuperscript{th} row and j\textsuperscript{th} column. 
M\textsubscript{ii}= the weighted diagonal element of the confusion matrix. 

It is mentioned above that we have particularly extracted MFCC and LPC features separately and concatenated them to get the final feature set. From MFCC and LPC methods we got 216 and 743 features respectively. Therefore, after concatenation, we had total features of 959. Usually using LPC feature extraction methods around 13 features are extracted, as we are implementing optimization techniques to get the best feature set, and it is quite evident that if the feature space is pretty vast then the performance of the entire model is improved by many folds.   
\begin{table}[htbp]
\centering
\caption{Quantitive evaluation results of KNN classifier using 5-fold validation on SAVEE dataset}
\label{table1}
\begin{tabular}{|l|l|l|l|l|}
\hline
Fold:  & Accuracy & Precision & Recall & F1 Score \\ \hline
Fold 1 & 96.56    & 97        & 97     & 97       \\ \hline
Fold 2 & 95.26    & 96        & 96     & 95       \\ \hline
Fold 3 & 97.81    & 98        & 98     & 98       \\ \hline
Fold 4 & 96.56    & 97        & 97     & 97       \\ \hline
Fold 5 & 96.56    & 97        & 97     & 97       \\ \hline
Mean \& STD: &
  \begin{tabular}[c]{@{}l@{}}96.55\%\\ ±0.80\end{tabular} &
  \begin{tabular}[c]{@{}l@{}}97.00\%\\ ±0.63\end{tabular} &
  \begin{tabular}[c]{@{}l@{}}97.00\%\\ ±0.63\end{tabular} &
  \begin{tabular}[c]{@{}l@{}}96.8\%\\ ±0.98\end{tabular} \\ \hline
\end{tabular}%

\end{table}

\begin{table}[htbp]
\centering
\caption{Quantitive evaluation results of MLP classifier using 5-fold validation on SAVEE dataset}
\label{table2}
\begin{tabular}{|l|l|l|l|l|}
\hline
Fold:  & Accuracy & Precision & Recall & F1 Score \\ \hline
Fold 1 & 97.91    & 98        & 98     & 98       \\ \hline
Fold 2 & 97.91    & 98        & 98     & 98       \\ \hline
Fold 3 & 96.88    & 97        & 97     & 97       \\ \hline
Fold 4 & 96.88    & 97        & 97     & 97       \\ \hline
Fold 5 & 97.91    & 98        & 98     & 98       \\ \hline
Mean \& STD: &
  \textbf{\begin{tabular}[c]{@{}l@{}}97.49\%\\ ±0.50\end{tabular}} &
  \textbf{\begin{tabular}[c]{@{}l@{}}97.6\%\\ ±0.48\end{tabular}} &
  \textbf{\begin{tabular}[c]{@{}l@{}}97.6\%\\ ±0.48\end{tabular}} &
  \textbf{\begin{tabular}[c]{@{}l@{}}97.6\%\\ ±0.48\end{tabular}} \\ \hline
\end{tabular}%

\end{table}

After concatenation the entire feature space becomes of size 480$\times$959, as we had 480 sample data and 959 features were extracted from each sample data. Manta Ray Optimization technique is used for optimized feature selection. It is obvious that the optimized feature space contains lesser features than that of the entire feature space, therefore the problem of over-fitting is already being taken care of by the optimization algorithm. 
\begin{figure}[htbp]
\centerline{\includegraphics[width=\columnwidth]{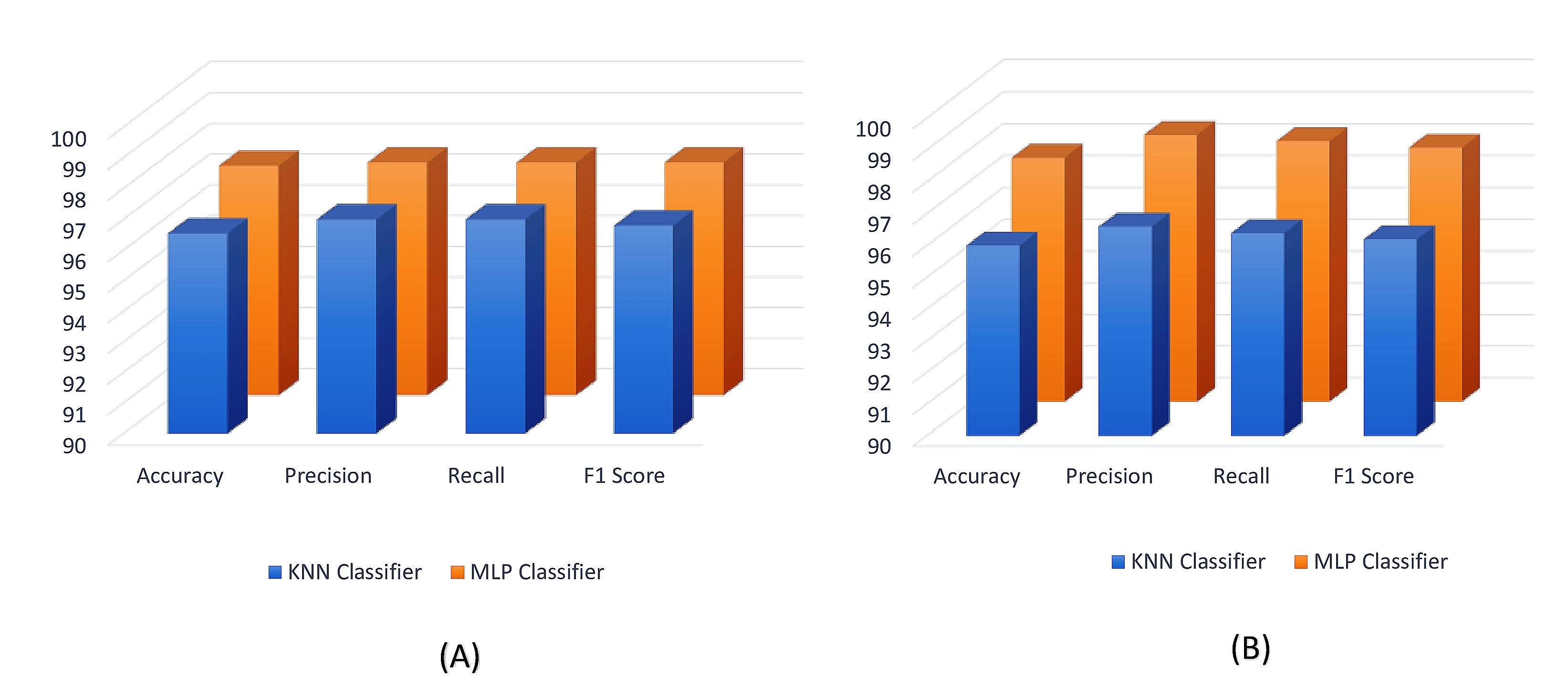}}
\caption{Combparative study of performance of classifiers (i.e. MLP and KNN) with same features on two diferrent datasets (A) Emo-DB dataset and (B) SAVEE dataset.}
\label{fig5}
\end{figure}
\begin{figure*}[htbp]
\centerline{\includegraphics[width=1.8\columnwidth]{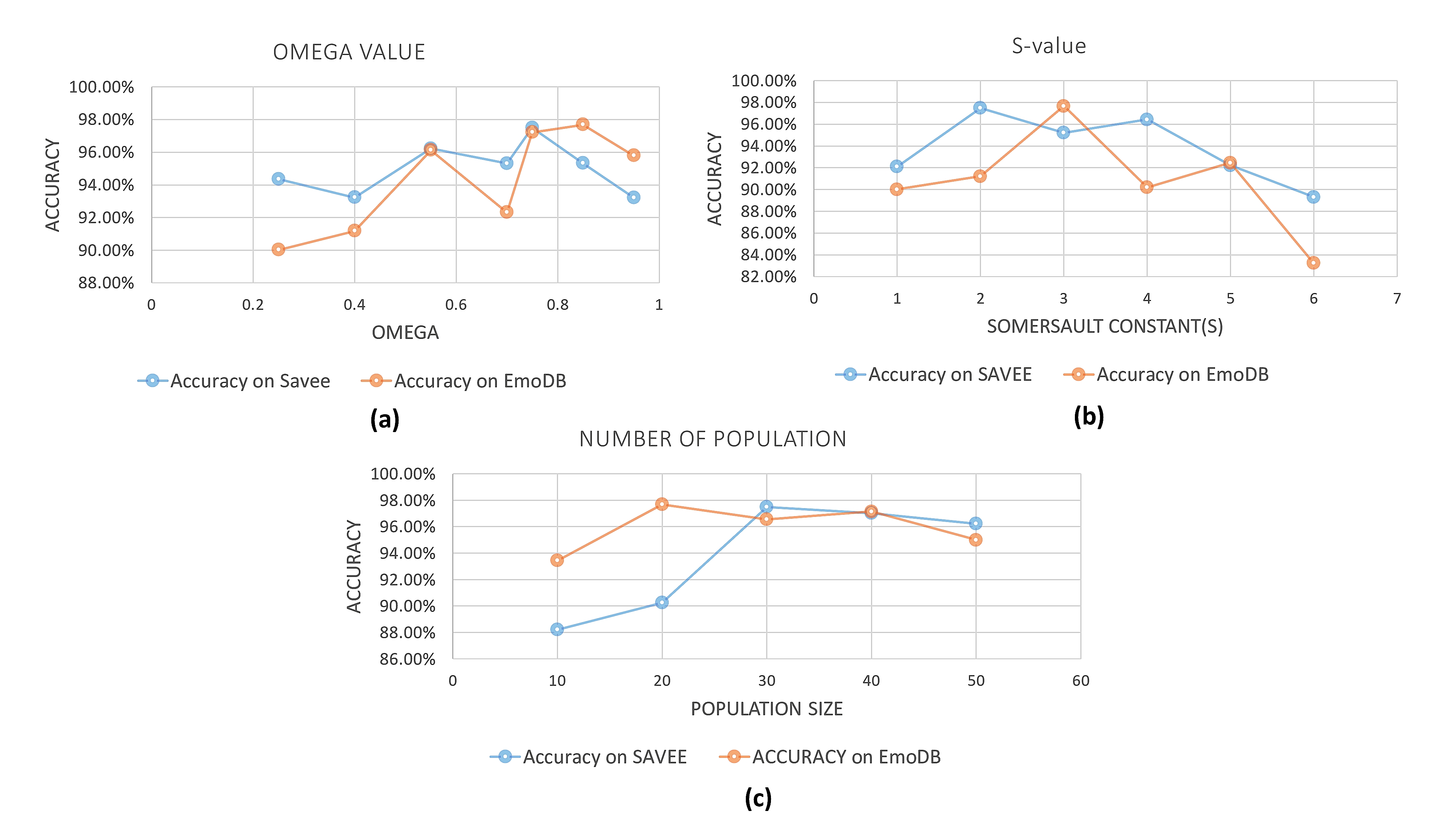}}
\caption{Combparative study of performance of classifiers (i.e. MLP and KNN) with same features on two diferrent datasets (A) Emo-DB dataset and (B) SAVEE dataset.}
\label{fig6}
\end{figure*}
Fig. 7 and Fig. 8 give the obtained ROC curves on both datasets with MLP and KNN calssifiers. With our proposed framework it is observed that the performance of MLP is better than that of the KNN. 
\begin{figure}[htbp]
\centerline{\includegraphics[width=\columnwidth]{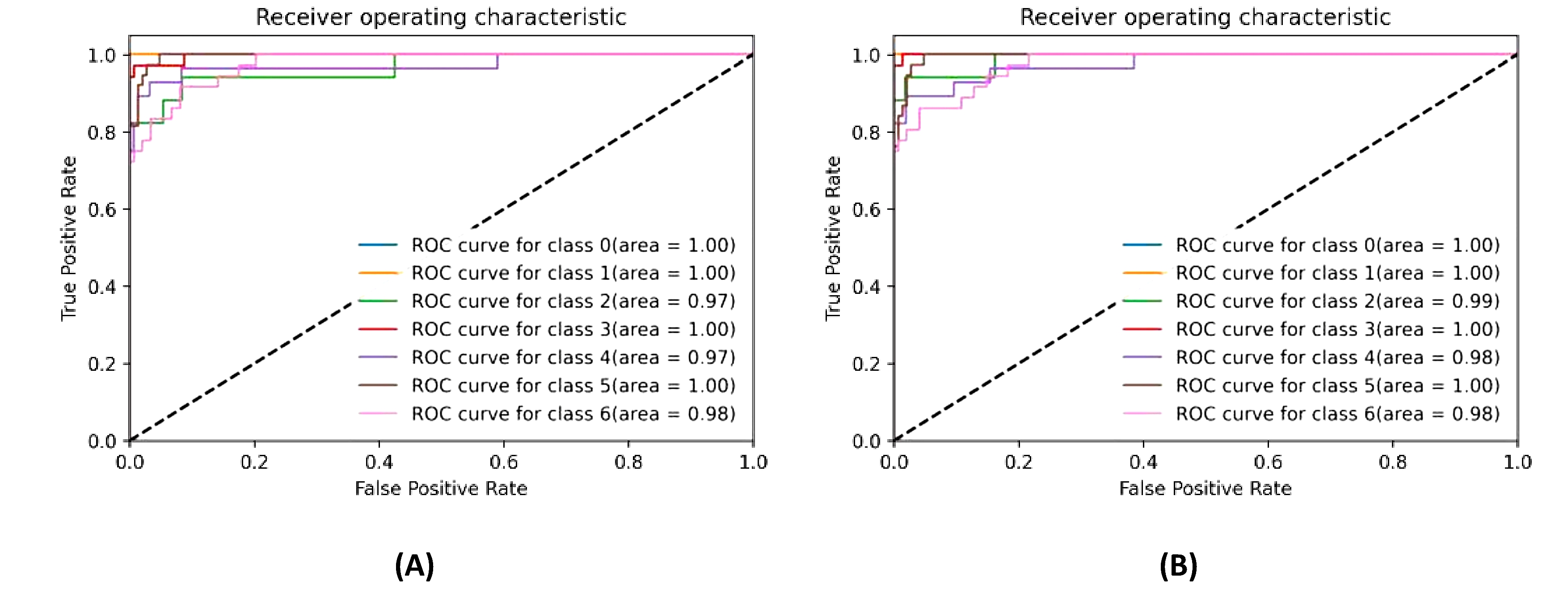}}
\caption{Reciever Operating Characteristics of KNN classifier on (A) Emo-DB dataset and (B) SAVEE dataset}
\label{fig9}

\centerline{\includegraphics[width=\columnwidth]{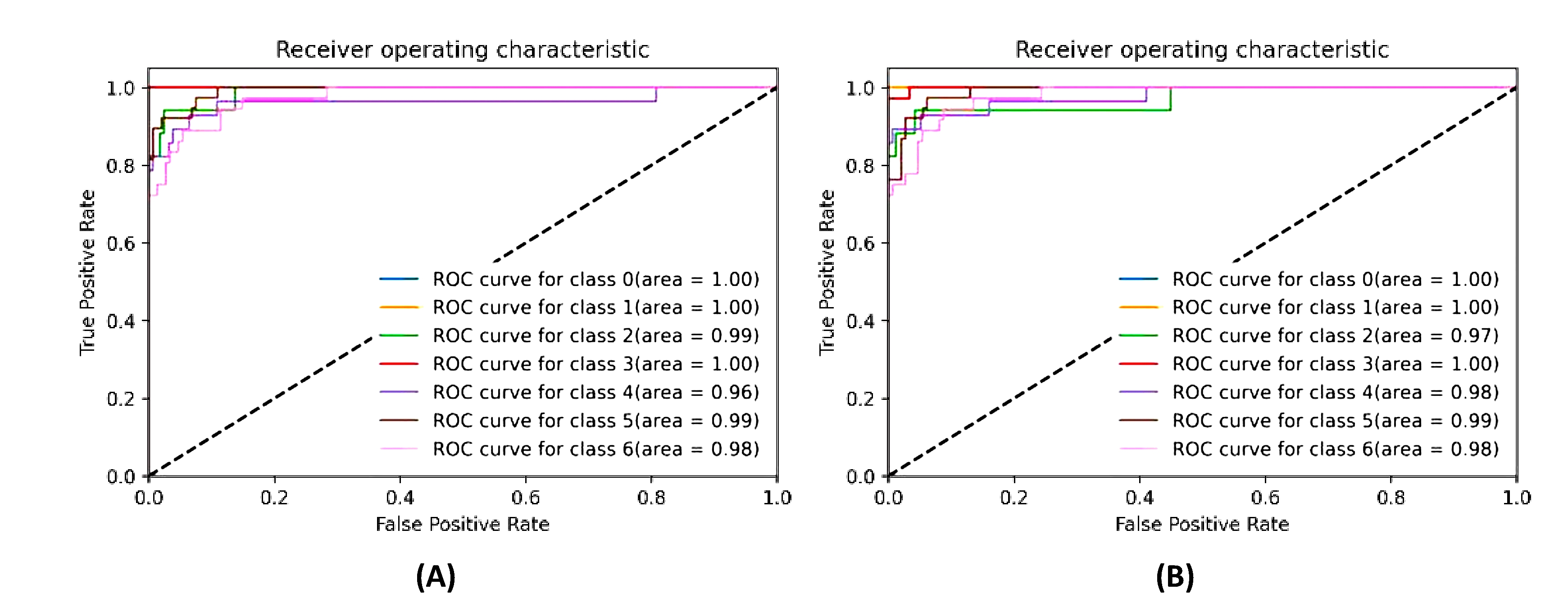}}
\caption{Reciever Operating Characteristics of MLP classifier on (A) Emo-DB dataset and (B) SAVEE dataset}

\end{figure}
\begin{table}[htbp]
\centering
\caption{Quantitive evaluation results of KNN classifier using 5-fold validation on Emo-DB dataset}
\label{table3}

\begin{tabular}{|l|l|l|l|l|}
\hline
Fold:  & Accuracy\% & Precision\% & Recall\% & F1 Score\% \\ \hline
Fold 1 & 96.00      & 97          & 96       & 96         \\ \hline
Fold 2 & 95.43      & 96          & 96       & 96         \\ \hline
Fold 3 & 95.96      & 96          & 96       & 96         \\ \hline
Fold 4 & 97.21      & 98          & 98       & 97         \\ \hline
Fold 5 & 95.43      & 96          & 96       & 96         \\ \hline
Mean \& STD: &
  \begin{tabular}[c]{@{}l@{}}96.006\%\\ ±0.65\end{tabular} &
  \begin{tabular}[c]{@{}l@{}}96.60\%\\ ±0.80\end{tabular} &
  \begin{tabular}[c]{@{}l@{}}96.40\%\\ ±0.80\end{tabular} &
  \begin{tabular}[c]{@{}l@{}}96.2\%\\ ±0.40\end{tabular} \\ \hline
\end{tabular}%

\end{table}

\begin{table}[htbp]
\centering
\caption{Quantitive evaluation results of MLP classifier using 5-fold validation on Emo-DB dataset}
\label{table4}

\begin{tabular}{|l|l|l|l|l|}
\hline
Fold:  & Accuracy\% & Precision\% & Recall\% & F1 Score\% \\ \hline
Fold 1 & 97.25      & 98          & 97       & 98         \\ \hline
Fold 2 & 98.31      & 99          & 99       & 98         \\ \hline
Fold 3 & 98.31      & 99          & 99       & 99         \\ \hline
Fold 4 & 98.31      & 99          & 99       & 99         \\ \hline
Fold 5 & 96.22      & 97          & 97       & 96         \\ \hline
Mean \& STD: &
  \textbf{\begin{tabular}[c]{@{}l@{}}97.68\%\\ ±0.83\end{tabular}} &
  \textbf{\begin{tabular}[c]{@{}l@{}}98.40\%\\ ±0.80\end{tabular}} &
  \textbf{\begin{tabular}[c]{@{}l@{}}98.20\%\\ ±0.97\end{tabular}} &
  \textbf{\begin{tabular}[c]{@{}l@{}}98.00\%\\ 1.09\end{tabular}} \\ \hline
\end{tabular}

\end{table}

To classify the dataset into given classes we have used KNN classifier with number of nearest neighborhood equals to 5 and MLP classifier consisting of two hidden layers, each having 5 neurons. Fig. 6 represents the comparison of the evaluation parameters using MLP and KNN classifiers on Emo-DB and SAVEE datasets. The bar graph shows that the MLP classifier performed better in both the cases. 
From Table II and Table IV we can say that our proposed model with MLP classifier achieves State-Of-The-Art accuracies of 97.06\% and 97.68\% on SAVEE and EmoDB datasets respectively. Apart from that KNN classifier also gives promising accuracies of 96.55\% and 96.006\% 7 class classification accuracies on aforementioned datasets.\\
In addition, to standardize our model performance we perform a comparative study of MantaRay with other popular optimization algorithms such as Genetic Algorithm [43], Particle Swarm [44] and Gray-Wolf Optimization algorithm\textbf[45]. The results of the optimization algorithms along with MantaRay is given by Table V. For SAVEE and EmoDB both datsets MantaRay achieves very good accuracies of 97.06\% and 97.68\% with significantly smaller feature space having only 43 and 61 features. This fact clearly supports the higher efficiency of our model.

\begin{table*}[]
\centering
\caption{Comparison of Manta Ray optimization with other feature selection algorithms in terms of accuracy, precision, recall, F1 score}
\label{tab7}
\resizebox{\textwidth}{!}{%
\begin{tabular}{|ccccccc|}
\hline
Dataset &
  \begin{tabular}[c]{@{}c@{}}Optimization   \\ Algorithm\end{tabular} &
  \begin{tabular}[c]{@{}c@{}}Number of selected   \\ features\end{tabular} &
  \begin{tabular}[c]{@{}c@{}}Accuracy \\ (rounded off)\end{tabular} &
  \begin{tabular}[c]{@{}c@{}}Precision\\ (rounded off)\end{tabular} &
  \begin{tabular}[c]{@{}c@{}}Recall \\ (rounded off)\end{tabular} &
  \begin{tabular}[c]{@{}c@{}}F1 score\\ (rounded off)\end{tabular} \\ \hline
\multirow{4}{*}{SAVEE}  & \begin{tabular}[c]{@{}c@{}}Genetic .\\ Algorithm\end{tabular}                 & 325 & 72.23\% & 73\% & 73\% & 74\% \\ \cline{2-7} 
                        & \begin{tabular}[c]{@{}c@{}}Grey Wolf   \\ Optimization Algorithm\end{tabular} & 89  & 84.31\% & 85\% & 87\% & 82\% \\ \cline{2-7} 
                        & \begin{tabular}[c]{@{}c@{}}Particle Swarm   \\ Optimization\end{tabular}      & 35  & 81.32\% & 81\% & 82\% & 83\% \\ \cline{2-7} 
                        & \begin{tabular}[c]{@{}c@{}}Manta Ray \\ Optimization\end{tabular}             & 43  & 97.06\% & 97\% & 98\% & 99\% \\ \hline
\multirow{4}{*}{Emo-DB} & \begin{tabular}[c]{@{}c@{}}Genetic .\\ Algorithm\end{tabular}                 & 230 & 85.91\% & 87\% & 86\% & 87\% \\ \cline{2-7} 
                        & \begin{tabular}[c]{@{}c@{}}Grey Wolf   \\ Optimization Algorithm\end{tabular} & 76  & 82.21\% & 83\% & 82\% & 83\% \\ \cline{2-7} 
                        & \begin{tabular}[c]{@{}c@{}}Particle Swarm   \\ Optimization\end{tabular}      & 26  & 87.29\% & 88\% & 87\% & 89\% \\ \cline{2-7} 
                        & \begin{tabular}[c]{@{}c@{}}Manta Ray \\ Optimization\end{tabular}             & 61  & 97.68\% & 99\% & 98\% & 98\% \\ \hline
\end{tabular}%
}
\end{table*}


The performances of the proposed model with KNN and MLP classifier on both SAVEE and EmoDB datasets are shown using bar-plots in Fig. 6.


Table VI and VII  display comparative analysis of the obtained result with the existing results so far. The results are selected for comparison in such a way that the experimentation parameters (e.g. train-test splitting ratio on a particular dataset) are more or less similar to maintain uniformity. It is observed that, in both the cases, our method outperforms the existing approaches by a good margin.

\section{Conclusion and future work}
This paper aims to contribute to the improvement of the speech emotion recognition task using a meta-heuristic Manta-Ray optimization algorithm for discarding redundant features and selecting the most accurate ones for classification. It implements two emotion recognition databases, two feature extraction methods, and achieves the best performance on both datasets. The neural network-based MLP classifier performed better in both the cases as compared to the KNN classifier. This is also justified from Fig. 5 and 6 where the area under ROC curves show better results in MLP classifier in both the datasets. However, we aim for validating our model on some other datasets, and also we would like to add some experimentation with different optimization algorithms. Further, we would like to assess the model performance with variations in sentence length as it has been found difficult to classify long sentences than small syllables earlier. Finally, we would like to perform experiments on some other languages too and try to evaluate the performance of native languages as there exists some relationship between accents and ease of classification.


\begin{thebibliography}{00}

\bibitem{}T. Bottesch and G. Palm, "Improving Classification Performance by Merging Distinct Feature Sets of Similar Quality Generated by Multiple Initializations of mRMR," 2015 IEEE Symposium Series on Computational Intelligence, Cape Town, 2015, pp. 328-334, doi: 10.1109/SSCI.2015.56. 

\bibitem{}S. Maniar and J. Shah, "Classification of Content based Medical Image Retrieval Using Texture and Shape feature with Neural Network", International Journal of Advances in Applied Sciences, vol. 6, no. 4, p. 368, 2017. Available: 10.11591/ijaas.v6.i4.pp368-374. 

\bibitem{}Sijun Qin, Jia Song, P. Zhang and Yue Tan, "Feature selection for text classification based on part of speech filter and synonym merge," 2015 12th International Conference on Fuzzy Systems and Knowledge Discovery (FSKD), Zhangjiajie, 2015, pp. 681-685, doi: 10.1109/FSKD.2015.7382024. 

\bibitem{}R. Zhang and Y. Yang, "Merging recovery feature network to faster RCNN for low-resolution images detection," 2017 IEEE Global Conference on Signal and Information Processing (GlobalSIP), Montreal, QC, 2017, pp. 1230-1234, doi: 10.1109/GlobalSIP.2017.8309157. 

\bibitem{}W. Zhao, Z. Zhang and L. Wang, "Manta ray foraging optimization: An effective bio-inspired optimizer for engineering applications", Engineering Applications of Artificial Intelligence, vol. 87, p. 103300, 2020. Available: 10.1016/j.engappai.2019.103300. 

 \bibitem{}J. Deng, Z. Zhang, E. Marchi and B. Schuller, "Sparse Autoencoder-Based Feature Transfer Learning for Speech Emotion Recognition," 2013 Humaine Association Conference on Affective Computing and Intelligent Interaction, Geneva, 2013, pp. 511-516, doi: 10.1109/ACII.2013.90. 

\bibitem{}R. Nakatsu, J. Nicholson and N. Tosa, "Emotion recognition and its application to computer agents with spontaneous interactive capabilities," 1999 IEEE Third Workshop on Multimedia Signal Processing (Cat. No.99TH8451), Copenhagen, Denmark, 1999, pp. 439-444, doi: 10.1109/MMSP.1999.793887. 
\bibitem{}S. Lalitha, A. Madhavan, B. Bhushan and S. Saketh, "Speech emotion recognition," 2014 International Conference on Advances in Electronics Computers and Communications, Bangalore, 2014, pp. 1-4, doi: 10.1109/ICAECC.2014.7002390.


\bibitem{}B. Schuller, S. Reiter, R. Muller, M. Al-Hames, M. Lang and G. Rigoll, "Speaker Independent Speech Emotion Recognition by Ensemble Classification," 2005 IEEE International Conference on Multimedia and Expo, Amsterdam, 2005, pp. 864-867, doi: 10.1109/ICME.2005.1521560. 

\bibitem{}J. Rong, G. Li and Y. Chen, "Acoustic feature selection for automatic emotion recognition from speech", Information Processing \& Management, vol. 45, no. 3, pp. 315-328, 2009. Available: 10.1016/j.ipm.2008.09.003. 

\bibitem{}K. Rao, S. Koolagudi and R. Vempada, "Emotion recognition from speech using global and local prosodic features", International Journal of Speech Technology, vol. 16, no. 2, pp. 143-160, 2012. Available: 10.1007/s10772-012-9172-2. 

\bibitem{}T. Nwe, S. Foo and L. De Silva, "Speech emotion recognition using hidden Markov models", Speech Communication, vol. 41, no. 4, pp. 603-623, 2003. Available: 10.1016/s0167-6393(03)00099-2. 

\bibitem{}S. Wu, T. Falk and W. Chan, "Automatic speech emotion recognition using modulation spectral features", Speech Communication, vol. 53, no. 5, pp. 768-785, 2011. Available: 10.1016/j.specom.2010.08.013. 

\bibitem{}H. Fayek, M. Lech and L. Cavedon, "Evaluating deep learning architectures for Speech Emotion Recognition", Neural Networks, vol. 92, pp. 60-68, 2017. Available: 10.1016/j.neunet.2017.02.013. 

\bibitem{}Y. Bengio, A. Courville and P. Vincent, "Representation Learning: A Review and New Perspectives", IEEE Transactions on Pattern Analysis and Machine Intelligence, vol. 35, no. 8, pp. 1798-1828, 2013. Available: 10.1109/tpami.2013.50. 

\bibitem{}A. Stuhlsatz, C. Meyer, F. Eyben, T. Zielke, G. Meier and B. Schuller, "Deep neural networks for acoustic emotion recognition: Raising the benchmarks," 2011 IEEE International Conference on Acoustics, Speech and Signal Processing (ICASSP), Prague, 2011, pp. 5688-5691, doi: 10.1109/ICASSP.2011.5947651. 

\bibitem{}J. Kim, "Bimodal Emotion Recognition using Speech and Physiological Changes", Robust Speech Recognition and Understanding, 2007. Available: 10.5772/4754 [Accessed 23 July 2020]. 

\bibitem{}Q. Mao, M. Dong, Z. Huang and Y. Zhan, "Learning Salient Features for Speech Emotion Recognition Using Convolutional Neural Networks," in IEEE Transactions on Multimedia, vol. 16, no. 8, pp. 2203-2213, Dec. 2014, doi: 10.1109/TMM.2014.2360798. 

\bibitem{}Z. Liu, Q. Xie, M. Wu, W. Cao, Y. Mei and J. Mao, "Speech emotion recognition based on an improved brain emotion learning model", Neurocomputing, vol. 309, pp. 145-156, 2018. Available: 10.1016/j.neucom.2018.05.005. 

\bibitem{}D. Nguyen, K. Nguyen, S. Sridharan, I. Abbasnejad, D. Dean and C. Fookes, "Meta Transfer Learning for Facial Emotion Recognition", 2018 24th International Conference on Pattern Recognition (ICPR), 2018. Available: 10.1109/icpr.2018.8545411 [Accessed 23 July 2020]. 

\bibitem{}N. Hajarolasvadi and H. Demirel, "3D CNN-Based Speech Emotion Recognition Using K-Means Clustering and Spectrograms", Entropy, vol. 21, no. 5, p. 479, 2019. Available: 10.3390/e21050479 [Accessed 23 July 2020]. 

\bibitem{}I. Pereira, D. Santos, A. Maciel and P. Barros, "Semi-supervised Model for Emotion Recognition in Speech", Artificial Neural Networks and Machine Learning – ICANN 2018, pp. 791-800, 2018. Available: 10.1007/978-3-030-01418-6\_77 [Accessed 23 July 2020]. 

\bibitem{}T. Danisman and A. Alpkocak, "Emotion Classification of Audio Signals Using Ensemble of Support Vector Machines", Lecture Notes in Computer Science, pp. 205-216. Available: 10.1007/978-3-540-69369-7\_23 [Accessed 23 July 2020]. 

\bibitem{}E. Albornoz, D. Milone and H. Rufiner, "Spoken emotion recognition using hierarchical classifiers", Computer Speech \& Language, vol. 25, no. 3, pp. 556-570, 2011. Available: 10.1016/j.csl.2010.10.001 [Accessed 23 July 2020]. 

 \bibitem{}P. Shen, Z. Changjun and X. Chen, "Automatic Speech Emotion Recognition using Support Vector Machine," Proceedings of 2011 International Conference on Electronic \& Mechanical Engineering and Information Technology, Harbin, 2011, pp. 621-625, doi: 10.1109/EMEIT.2011.6023178. 

\bibitem{}K. Wang, N. An, B. N. Li, Y. Zhang and L. Li, "Speech Emotion Recognition Using Fourier Parameters," in IEEE Transactions on Affective Computing, vol. 6, no. 1, pp. 69-75, 1 Jan.-March 2015, doi: 10.1109/TAFFC.2015.2392101. 

\bibitem{}C. Wu and W. Liang, "Emotion Recognition of Affective Speech Based on Multiple Classifiers Using Acoustic-Prosodic Information and Semantic Labels," in IEEE Transactions on Affective Computing, vol. 2, no. 1, pp. 10-21, Jan.-June 2011, doi: 10.1109/T-AFFC.2010.16.
\bibitem{}B. Schuller, G. Rigoll and M. Lang, "Hidden Markov model-based speech emotion recognition," 2003 IEEE International Conference on Acoustics, Speech, and Signal Processing, 2003. Proceedings. (ICASSP '03)., Hong Kong, 2003, pp. II-1, doi: 10.1109/ICASSP.2003.1202279. 
\bibitem{} X. Huahu, G. Jue and Y. Jian, "Application of Speech Emotion Recognition in Intelligent Household Robot," 2010 International Conference on Artificial Intelligence and Computational Intelligence, Sanya, 2010, pp. 537-541, DOI: 10.1109/AICI.2010.118. 

\bibitem{}Schuller, B., 2018. Speech emotion recognition. Communications of the ACM, 61(5), pp.90-99. 

\bibitem{}M. Callejas-Cuervo, L. Martínez-Tejada and A. Alarcón-Aldana, "Emotion recognition techniques using physiological signals and video games –Systematic review–", Revista Facultad de Ingeniería, vol. 26, no. 46, 2017. Available: 10.19053/01211129.v26.n46.2017.7310. 

\bibitem{}L. Low, M. Maddage, M. Lech, L. Sheeber and N. Allen, "Detection of Clinical Depression in Adolescents’ Speech During Family Interactions", IEEE Transactions on Biomedical Engineering, vol. 58, no. 3, pp. 574-586, 2011. Available: 10.1109/tbme.2010.2091640. 

 \bibitem{}Liu ZT, Wu M, Cao WH, Mao JW, Xu JP, Tan GZ. Speech emotion recognition based on feature selection and extreme learning machine decision tree. Neurocomputing. 2018 Jan 17;273:271-80.

\bibitem{}Ververidis D, Kotropoulos C. Fast and accurate sequential floating forward feature selection with the Bayes classifier applied to speech emotion recognition. Signal processing. 2008 Dec 1;88(12):2956-70.
\bibitem{}Özseven T. A novel feature selection method for speech emotion recognition. Applied Acoustics. 2019 Mar 1;146:320-6.
\bibitem{}Sheikhan M, Bejani M, Gharavian D. Modular neural-SVM scheme for speech emotion recognition using ANOVA feature selection method. Neural Computing and Applications. 2013 Jul 1;23(1):215-27.
\bibitem{}Gharavian D, Sheikhan M, Nazerieh A, Garoucy S. Speech emotion recognition using FCBF feature selection method and GA-optimized fuzzy ARTMAP neural network. Neural Computing and Applications. 2012 Nov 1;21(8):2115-26.

\bibitem{}Muthusamy H, Polat K, Yaacob S. Particle swarm optimization based feature enhancement and feature selection for improved emotion recognition in speech and glottal signals. PloS one. 2015 Mar 23;10(3):e0120344.
\bibitem{}Logan B. Mel frequency cepstral coefficients for music modeling. InIsmir 2000 Oct 23 (Vol. 270, pp. 1-11).
\bibitem{}Atal BS, Hanauer SL. Speech analysis and synthesis by linear prediction of the speech wave. The journal of the acoustical society of America. 1971 Aug;50(2B):637-55.
\bibitem{}Atal BS. Effectiveness of linear prediction characteristics of the speech wave for automatic speaker identification and verification. the Journal of the Acoustical Society of America. 1974 Jun;55(6):1304-12.

\bibitem{}Hermansky H. Perceptual linear predictive (PLP) analysis of speech. the Journal of the Acoustical Society of America. 1990 Apr;87(4):1738-52.
\bibitem{} Holland JH. Genetic algorithms. Scientific american. 1992 Jul 1;267(1):66-73.
\bibitem{}Kennedy J, Eberhart R. Particle swarm optimization. InProceedings of ICNN'95-International Conference on Neural Networks 1995 Nov 27 (Vol. 4, pp. 1942-1948). IEEE.
\bibitem{} Emary E, Zawbaa HM, Hassanien AE. Binary grey wolf optimization approaches for feature selection. Neurocomputing. 2016 Jan 8;172:371-81.

\end{thebibliography}
\end{document}